\documentclass[aps,prb,twocolumn,floatfix,superscriptaddress,showpacs,showkeys]{revtex4}
\usepackage{amssymb}
\usepackage{amsmath}
\usepackage{graphicx}

\begin{document}

\title{Magnetoresistance of nondegenerate quantum electron channels
formed on the surface of superfluid helium}
\author{Yu.P.~Monarkha}
\affiliation{Institute for Low Temperature Physics and
Engineering, 47 Lenin Avenue, 61103, Kharkov, Ukraine}
\author{S.S.~Sokolov}
\affiliation{Institute for Low Temperature Physics
and Engineering, 47 Lenin Avenue, 61103, Kharkov, Ukraine}
\affiliation {Departamento de F\'{i}sica, Universidade Federal de
S\~{a}o Carlos, 13565-905 S\~{a}o Carlos, S\~{a}o Paulo, Brazil}
\author{G.-Q.~Hai}
\affiliation{$^3$Instituto de F\'{i}sica de S\~{a}o Carlos,
Universidade de S\~{a}o Paulo, 13560-970 Sao Carlos, Brazil}
\author{Nelson Studart}
\affiliation {Departamento de F\'{i}sica, Universidade Federal de
S\~{a}o Carlos, 13565-905 S\~{a}o Carlos, S\~{a}o Paulo, Brazil}


\begin{abstract}
Transport properties of quasi-one-dimensional nondegenerate quantum wires
formed on the surface of liquid helium in the presence of a normal magnetic
field are studied using the momentum balance equation method and the memory
function formalism. The interaction with both kinds of scatterers available
(vapor atoms and capillary wave quanta) is considered. We show that unlike
classical wires, quantum nondegenerate channels exhibit strong
magnetoresistance which increases with lowering the temperature.
\end{abstract}

\maketitle

\section{Introduction}

Electrons trapped on the free surface of liquid helium form a
nondegenerate two-dimensional electron system (2DES) ~\cite{And97}
whose properties are complementary to the properties of the
degenerate 2DES created in semiconductor structures. Considerable
experimental and theoretical research has been performed on the
quantum magnetotransport in such an almost pure
and highly correlated 2DES (for a recent review, see Ref.~\ \cite%
{MonTesWyd02}). In the presence of a strong magnetic field applied
in the normal direction to the system the electron energy spectrum
is squeezed into the set of Landau levels slightly broadened due
to the interaction with scatterers. For surface electrons (SE) on
liquid helium, the broadening of Landau levels is extremely
narrow, usually much smaller than temperature, which is the origin
of the unconventional Hall effect observed in this
system under different experimental conditions~\cite%
{PetSchLea94,LeaFozRic94,MonTesWyd99}.

The SE can be confined in quasi one-dimensional (1D) channels,
when the helium surface is curved by capillary forces in the
presence of a specially constructed dielectric
substrate~\cite{KirMonKov93,ValHei98}. The conducting channels are
formed in the valleys of the helium relief because of the strong
holding electric field $E_{\bot }$ applied normally to the
surface. These channels can be considered as the nondegenerate
version of quantum wires created in semiconductor structures. In
the presence of a
strong normal magnetic field, the electron states in the channels~\cite%
{SokStu95} resemble the current-carrying edge states of the
quantum Hall effect systems~\cite{Gir98}. There is also an
interesting evidence for the
self-organized current filaments in the helium microchannels~\cite%
{GlaDotFoz01}.

\begin{figure}[tbph]
\includegraphics*[width=0.8\linewidth]{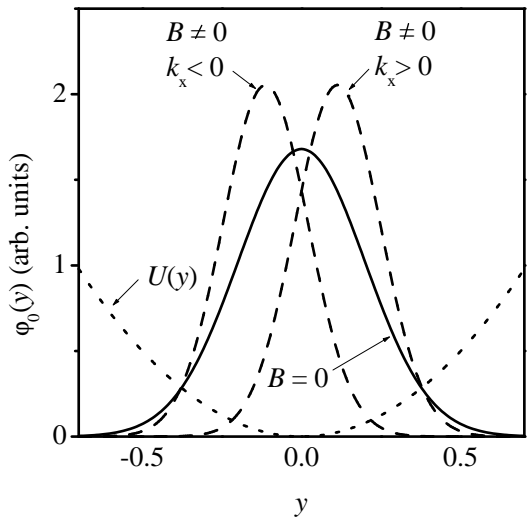}
\caption{Profiles
of the electron wave function across the channel for $B=0$
(\textit{solid line}) and for a characteristic $B\neq 0$
(\textit{dashed lines}). The confining potential $U(y)$ is
indicated by the \textit{dotted line}.} \label{f1}
\end{figure}

The confining potential affects crucially the energy spectrum of electrons
subject to the magnetic field $B$, because it removes the degeneracy of
Landau levels. For example, in 2DES the Landau spectrum does not depend on
the orbit center coordinate $Y=-l_{B}^{2}k_{x}$, and it is purely discrete $%
\varepsilon _{n,k_{x}}=\hbar \omega _{c}(n+1/2)$, where $\hbar k_{x}$ is the
electron momentum along the $x$-direction, $n=0,1,2...$, $l_{B}=(\hbar
c/eB)^{1/2}$ is the magnetic length and $\omega _{c}=eB/m_{e}c$ is the
cyclotron frequency defined in terms of the free electron mass $m_{e}$. In
contrast, for the parabolic confining potential $U(y)=m_{e}\omega
_{0}^{2}y^{2}/2$, the electron energy spectrum under the magnetic field has
a continuous term depending on the electron momentum along the channel~\cite%
{SokStu95}:
\begin{equation}
\varepsilon _{n,k_{x}}=\frac{\hbar ^{2}k_{x}^{2}}{2m_{B}}+\hbar
\Omega \left( n+\frac{1}{2}\right) ,  \label{e1}
\end{equation}%
where
\begin{equation}
m_{B}=m_{e}\left( 1+\frac{\omega _{c}^{2}}{\omega _{0}^{2}}\right)
,\,\,\Omega =\sqrt{\omega _{0}^{2}+\omega _{c}^{2}}.  \label{e2}
\end{equation}%
The orbit center coordinate $Y_{k_{x}}=-(\omega _{c}/\Omega )l_{y}^{2}k_{x}$
(here $l_{y}^{2}=\hbar /m_{e}\Omega $ is the typical electron localization
length across the channel) is determined by the interplay of the magnetic
field and the confining potential. Thus, the magnetic field shifts the
electron wave function in the channel $\varphi _{n}(y-Y_{k_{x}})$ to the
left or the right depending on the sign of $k_{x}$, as shown in Fig.~\ref{f1}%
. It increases also the frequency of the discrete part of the
electron spectrum ($\omega _{0}\rightarrow \Omega $) and the
effective mass of charge carriers ($m_{e}\rightarrow m_{B}\propto
\Omega ^{2}$). Such an unusual behavior of the electron effective
mass is very important for channel magnetotransport, localization
effects, and the polaronic transition (for a review on the last
two topics, see~\cite{StuSok97}).

In this work we report the theory of quantum magnetotransport in
Q1D channels for highly correlated electrons, when the
electron-electron collision rate $\nu _{ee}$ is much higher than
the effective collision frequency $\nu $ due to available
scatterers. In this regime, usually realized for SE on helium, the
memory function formulation for the electron
conductivity~\cite{GotHaj78} and the momentum balance equation
approach leads to the same conductivity equation. The important
advantage of these methods is that the electron conductivity along
the channel can be quite generally expressed in terms of the
electron dynamical structure factor (DSF) $S(\mathbf{q},\omega )$,
which allows to track the origin of the strong dependence of the
electron mobility on the magnetic field. We find that, in contrast
with classical wires, the resistivity of the nondegenerate quantum
wires formed on the surface of liquid helium is strongly affected
by the magnetic field. This effect depends strongly on the
temperature. The results obtained here are compared with that
found for the pure 2D case.

\section{Basic relations}

\subsection{Electron channel states}

In the presence of a strong holding electric field $E_{\bot }$ directed
along the $z$-axis the electrons in the channel are gathered near the
minimum of the channel profile which we describe by the semicircular form $%
z(y)=R(1-\sqrt{1-y^{2}/R^{2}})\simeq y^{2}/2R$ if $y\ll R$ (here $R$ is the
curvature radius of the helium surface which usually ranges from $10^{-3}$
to $10^{-4}$ cm). As a result, the potential energy of an electron across
the channel can be approximated by the parabolic potential $
U(y)=m_{e}\omega _{0}^{2}y^{2}/2$, where $\omega _{0}^{2}=eE_{\perp }/m_{e}R$%
. As the magnetic filed $B$ is applied in the direction parallel to the
holding electric field, we use the Landau gauge for the vector potential $%
\mathbf{A} =(-By,0,0)$ to take the advantage of the translational invariance
along the channel ($x$-axis). The electron motion along the $y$ direction is
quantized and the electron energy spectrum has the form given by Eq.~(\ref%
{e1}).

Because the curvature radius is much larger than $l_{y}$, we restrict our
study to the model of a 2DES subject to the oscillatory confining potential $%
U(y)$. The electron wave function in the channel is defined as

\begin{equation}
\psi _{n,k_{x}}(x,y)=\frac{1}{\sqrt{L_{x}}}\exp (\mathrm{i}k_{x}x)\varphi
_{n}[(y-Y_{k_{x}})/l_{y}],  \label{e3}
\end{equation}
where $\varphi _{n}(x)$ is the Hermite functions, $L_{x}$ is the linear
dimension of the channel and, as stated above, $Y_{k_{x}}=-(\omega
_{c}/\Omega )l_{y}^{2}k_{x}$.

In general, we have to include the interaction of the electron spin with the
magnetic field. This leads to the energy spectrum
\[
\varepsilon _{n,k_{x}}=\frac{\hbar ^{2}k_{x}^{2}}{2m_{B}}+\hbar
\Omega \left( n+\frac{1}{2}\right) +\hbar \omega _{c}\sigma ,
\]
where $\sigma =\pm 1/2$ is the spin projection eigenvalue. In the pure 2DES (%
$\Omega =\omega _{c}$), the inclusion of spin causes an additional
degeneracy of the electron states: the energy levels with $n+1$, $\sigma =1/2
$ and $n$, $\sigma =-1/2$ coincide. For electrons in channels, these levels
are split because $\Omega >\omega _{c}$.

The magnetic field does not modify the nature of the electron motion along
the channel, and the continuous part of the electron spectrum has the usual
form $\varepsilon _{k_{x}}=\hbar ^{2}k_{x}^{2}/2m_{B}$. Still, the magnetic
field increases strongly the effective mass $m_{B}$, and, for electrons with
a fixed density (as it is indeed for SE on helium), it reduces the Fermi
energy. At low electron densities we can consider only the lowest level with
$n=0$. Then, in the absence of the magnetic field, the Fermi momentum $k_{%
\mathrm{F}}=\pi n_{\mathrm{ch}}/2$ (here $n_{\mathrm{ch}}$ is the linear
electron density) and the 1D Fermi energy $\varepsilon _{\mathrm{F}}\equiv
\varepsilon _{0}$, where $\varepsilon _{0}=\pi ^{2}\hbar ^{2}n_{\mathrm{ch}%
}^{2}/8m_{e}$. The magnetic field splits the lowest level, and for a fixed
channel density, the 1D Fermi energy and the total energy of an electron at
the Fermi level depend strongly on the energy parameter $\varepsilon
_{B}=\pi ^{2}\hbar ^{2}n_{\mathrm{ch}}^{2}/8m_{B}$ which decreases with $B$
because of the mass enhancement. The total energy at the Fermi-level can be
written as
\begin{eqnarray}
\varepsilon _{\mathrm{total}} &=&4\varepsilon _{B}+\hbar \left( \Omega
-\omega _{c}\right) /2,\mathrm{\ if\ }\hbar \omega _{c}>4\varepsilon _{B},
\label{e4} \\
&=&\varepsilon _{B}+\hbar \Omega /2+\hbar ^{2}\omega _{c}^{2}/16\varepsilon
_{B},\mathrm{\ if\ }\hbar \omega _{c}<4\varepsilon _{B}.  \nonumber
\end{eqnarray}
It should be noted that the field dependence of $\varepsilon _{\mathrm{total}%
}$ is nonmonotonous: the total energy defined above increases at low fields (%
$\hbar \omega _{c}<4\varepsilon _{B}$), and decreases steadily at high
fields ($\hbar \omega _{c}>4\varepsilon _{B}$). The decrease of the Fermi
energy and the total energy with $B$ means that for finite temperatures and
in the limit of high fields the electron channel with a fixed density
eventually becomes a nondegenerate system. Because the interaction with
scatterers does not involve the electron spin, we shall disregard it when
calculating the scattering matrix elements.

\subsection{Dynamical structure factor}

As stated in the Introduction, the equilibrium electron DSF $S(\mathbf{q}%
,\omega )$ plays an important role in the quantum transport theory of highly
correlated electrons. We define it as
\begin{equation}
S(\mathbf{q},\omega )=\frac{1}{N_{e}}\int\limits_{-\infty }^{\infty }e^{%
\mathrm{i}\omega t}\left\langle n_{\mathbf{q}}(t)n_{-\mathbf{q}%
}(0)\right\rangle dt,  \label{e5}
\end{equation}%
where $n_{\mathbf{q}}=\sum_{e}\exp (-i\mathbf{q}\cdot \mathbf{r})$
is
the density fluctuation operator, $\mathbf{q}$ is the 2D wave vector and $%
\mathbf{r}$ is the 2D position of an electron. For the energy
spectrum given by Eq.~(\ref{e1}), we can evaluate the average
$\left\langle ...\right\rangle $ straightforwardly to obtain
\begin{eqnarray}
S(\mathbf{q},\omega )=\frac{2\pi }{N_{e}}\sum_{k_{x},k_{x}^{\prime
}}\sum_{n,n^{\prime }}f(\varepsilon _{n,k_{x}})\left[
1-f(\varepsilon _{n^{\prime },k_{x}^{\prime }})\right] \nonumber \\
\times \left| \left\langle n,k_{x}\right|
e^{-\mathrm{i}\mathbf{q}\cdot \mathbf{r}}\left| n^{\prime
},k_{x}^{\prime }\right\rangle \right| ^{2}\delta (\varepsilon
_{n,k_{x}}-\varepsilon _{n^{\prime },k_{x}^{\prime }}+\hbar \omega
),
  \label{e6}
\end{eqnarray}
where $f(\varepsilon )$ is the Fermi distribution function.

The SE on liquid helium form usually a nondegenerate system. Therefore we
can disregard $f(\varepsilon _{n^{\prime },k_{x}^{\prime }})$ as compared to
1 in Eq. (\ref{e6}). Then, introducing the notation
\begin{equation}
J_{n,n^{\prime }}(q_{x},q_{y})=\int\limits_{-\infty }^{\infty }e^{\mathrm{i}
q_{y}l_{y}\tau }\varphi _{n}(\tau )\varphi _{n^{\prime }}(\tau
-q_{x}l_{y}\omega _{c}/\Omega )d\tau ,  \label{e7}
\end{equation}
the DSF can be found in the form
\begin{eqnarray}
S(\mathbf{q},\omega )=\hbar \left( 1-e^{-\hbar \Omega /T}\right) \sqrt{\frac{%
\pi }{\varepsilon _{q_{x}}T}}\sum_{n,n^{\prime }}e^{-\hbar \Omega n/T}\left|
J_{n,n^{\prime }}\right| ^{2}  \nonumber \\
\times \exp \left\{ -\frac{\left[ \varepsilon _{q_{x}}-\hbar \omega -\hbar
\Omega (n-n^{\prime })\right] ^{2}}{4\varepsilon _{q_{x}}T}\right\} \hspace{%
1cm}  \label{e8}
\end{eqnarray}
with
\[
\left| J_{n,n^{\prime }}(x_{\mathbf{q}})\right| ^{2}=\frac{\min (n,n^{\prime
})!}{\max (n,n^{\prime })!}e^{-x_{\mathbf{q}}}x_{\mathbf{q}}^{\left|
n^{\prime }-n\right| }\left[ L_{\min (n,n^{\prime })}^{\left| n^{\prime
}-n\right| }(x_{\mathbf{q}})\right] ^{2},
\]
and
\[
x_{\mathbf{q}}=\left( q_{y}^{2}+q_{x}^{2}\frac{\omega _{c}^{2}}{\Omega ^{2}}
\right) \frac{l_{y}^{2}}{2}.
\]
Here $L_{n}^{m }(x)$ are the associated Laguerre polynomials.

The matrix elements $\left| J_{n,n^{\prime }}(x_{\mathbf{q}})\right| ^{2}$
restrict differently the wave numbers $q_{x}$ and $q_{y}$. For the lowest
level $n=n^{\prime }=0$, we have
\begin{equation}
\left| J_{0,0}(q_{x},q_{y})\right| ^{2}=\exp \left( -\frac{%
q_{y}^{2}l_{y}^{2} }{2}-\frac{q_{x}^{2}l_{y}^{2}}{2}\frac{\omega _{c}^{2}}{%
\Omega ^{2}}\right) .  \label{e9}
\end{equation}
In the extreme case $B=0$ this equation restricts only transverse
wave numbers $q_{y}$ owing to the channel confining potential. For
high
magnetic fields $\Omega \simeq \omega _{c}$ both wave numbers $q_{x}$ and $%
q_{y}$ enter into these matrix elements in the same way, which is usual for
the quantum magnetotransport in the 2DES. An additional restriction on $q_{x}
$ appears because of the factor $\exp (-\varepsilon _{q_{x}}/4T)$, but it
becomes less important for strong fields when $m_{B}\gg m_{e}$.

The channel DSF $S(\mathbf{q},\omega )$ exhibit features which are
typical for both the free electron gas under zero magnetic field
and the 2DES in the presence of a strong perpendicular magnetic
field. We first pay attention to the singularity
$S(\mathbf{q},\omega )\propto 1/\left| q_{x}\right| $ and the
factor $\exp (-\varepsilon _{q_{x}}/4T)$ which are inherent for
free electrons. They concern the $q_{x}$-component only,
reflecting the free electron motion along the channel. In the
ultra-quantum limit $T\ll \hbar \Omega$, Eq.~(\ref{e8}) can be
approximated by the terms with $n=0$. Then the channel DSF is
given as a sum of Gaussian terms broadened by the width parameter
$\Gamma ^{\ast }=2\sqrt{\varepsilon _{q_{x}}T}$, which exhibit the
resonant behavior as $\omega -n^{\prime }\Omega\rightarrow 0$.
This result is similar to that obtained for the 2D Coulomb liquid
under a normal magnetic field~\cite{MonTesWyd02}. In the limiting
case $\Gamma ^{\ast }\rightarrow 0$, the DSF is a sum of
delta-functions reflecting the singular nature of the 2DES under
the magnetic field $B$~\cite{AndFowSte82}.

\begin{figure}[tbp]
\includegraphics*[width=0.8\linewidth]{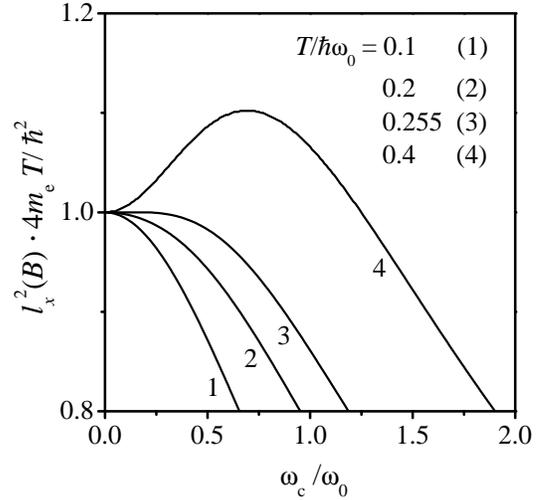}
\caption{The magnetic field dependence of normalized $l_{x}^{2}$
which determines the momentum exchange along the channel in scattering events $%
q_{x}\sim 1/l_{x}$. }
\label{f2}
\end{figure}

The broadening parameter of the Gaussians in Eq.~(\ref{e8}) $\Gamma ^{\ast
}=2\sqrt{\varepsilon _{q_{x}}T}$ can be estimated combining two exponents
proportional to $q_{x}^{2}$ by
\[
\frac{q_{x}^{2}l_{y}^{2}}{2}\frac{\omega _{c}^{2}}{\Omega ^{2}}+\frac{%
\varepsilon _{q_{x}}}{4T}\equiv \frac{q_{x}^{2}l_{x}^{2}}{2},
\]
where we have defined
\begin{equation}
l_{x}^{2}=l_{y}^{2}\frac{\omega _{c}^{2}}{\Omega ^{2}}+\frac{\hbar ^{2}}{%
4m_{B}T}.  \label{e9b}
\end{equation}
Then the typical electron wave numbers $q_{x}\sim \sqrt{2}/l_{x}$. In
general $l_{x}^{2}(B)$ is a nonmonotonous function of the magnetic field, as
shown in Fig.~\ref{f2}, where we have used the dimensionless parameters $%
T/\hbar \omega _{0}$ and $\omega _{c}/\omega _{0}$. We shall see that this
behavior affects the channel magnetotransport.

At low temperatures ($T/\hbar \omega _{0}<0.255$), $l_{x}$ decreases
steadily with $B$, and when $l_{x}\approx l_{y}\omega _{c}/\Omega $ we have $%
\Gamma ^{\ast }\sim 2\sqrt{T\hbar \Omega }\omega _{0}/\omega _{c}\equiv
\sqrt{2}eE_{\mathrm{ch}}l_{y}$, where $eE_{\mathrm{ch}}=\sqrt{2m_{e}T}\Omega
\omega _{0}/\omega _{c}$, with $E_{\mathrm{ch}}$ being a characteristic
electric field. For strong fields $\Omega \simeq \omega _{c}$, $eE_{\mathrm{%
ch} }\simeq \omega _{0}\sqrt{2m_{e}T}$ is the force acting on an
electron at $y=\sqrt{2T/m_{e}\omega _{0}^{2}}$ due to the
confining potential. It is interesting that in this limiting case,
the temperature and field dependencies of the broadening of the
channel DSF $\Gamma ^{\ast }\propto \sqrt{T/B}$ are the same as
that induced by strong internal forces~in the 2D Coulomb
liquid~\cite{MonTesWyd02}. Therefore, the usual many-electron
effects on the channel magnetotransport can be disregarded, if the
average
internal electric field of the fluctuational origin $E_{\mathrm{f}}\approx 3%
\sqrt{T} n_{s}^{3/2}$~\cite{LeaFozKri97} is smaller than the channel field $%
E_{\mathrm{ch}}\simeq \omega _{0}\sqrt{2m_{e}T}/e$.

\subsection{Channel magnetotransport}

In this work we primarily consider the DC magnetotransport, which means that
the frequency of the driving electric field is zero or negligibly low.
Therefore, the current across the channel is assumed to be zero ($J_{y}=0$
and the Lorentz force is balanced by the field of the confining potential).
Then for highly correlated electrons, the electrical current along the wire $%
J_{x}$ can be easily found by balancing the force of the driving
electric field $N_{e}eE_{x}$ and the frictional force
$F_{\mathrm{scat}}$ acting on the electron system due to the
scatterers ~\cite{MonTesWyd02}. In the linear transport regime the
absolute value of the frictional force is
proportional to the current or the average electron velocity in the channel $%
u_{\mathrm{\ av }}$, and can be written as $F_{\mathrm{scat}}=-N_{e}m_{e}\nu
_{\mathrm{\ eff} }(B)u_{\mathrm{av}}$. The proportionality factor $\nu _{%
\mathrm{eff} }(B) $ is called the effective collision frequency, which can
be found by evaluating the momentum loss of the electron system per unit
time. Then the current along the channel is given by
\begin{equation}
J_{x}\equiv eN_{e}u_{\mathrm{av}}=N_{e}e\mu E_{x},\,\,\mu =e/(m_{e}\nu _{%
\mathrm{eff}}),  \label{e10}
\end{equation}
where $\mu $ is the channel mobility. It is well known that for classical
thin wires $\nu _{\mathrm{eff}}(B)$ does not depend on $B$ (it coincides
with the conventional collision rate) and the theory gives zero
magnetoresistance. In the following we shall see that there is a strong
magnetoresistance for nondegenerate quantum wires.

In order to find $F_{\mathrm{scat}}$ we can calculate the momentum absorbed
by scatterers per unit time $F_{\mathrm{scat}}=-\overset{\cdot }{P}$. For SE
on helium, the only scatterers are helium vapor atoms and capillary wave
quanta (ripplons). At low temperatures $T\lesssim 0.5\,\mathrm{K}$ the
electrons are scattered predominantly by ripplons, because the vapor atom
density decreases with cooling at an exponential rate. In this case the
interaction Hamiltonian is proportional to the electron density fluctuation
operator $n_{\mathbf{q}}$ and given by
\[
H_{\mathrm{int}}=\frac{1}{\sqrt{S_{\mathrm{A}}}}\sum_{\mathbf{q}}V_{q}\xi _{%
\mathbf{q}}n_{-\mathbf{q}},
\]%
where $S_{\mathrm{A}}$ is the surface area, $V_{q}$ is the
electron-ripplon coupling, and $\xi _{\mathbf{q}}$ is the
Fourier-component of the surface-displacement operator $\xi
(\mathbf{r})$. Then, following Refs. ~\
\cite{VilMon89,MonTesWyd02}, the frictional force can be quite
generally expressed in terms of the electron dynamical structure
factor
\begin{eqnarray}
F_{\mathrm{scat}} &=&\frac{N_{e}}{\hbar S_{\mathrm{A}}}\sum_{\mathbf{q}%
}q_{x}V_{q}^{2}Q_{q}^{2}\hspace{1cm}  \nonumber \\
&&\times \left[ N_{\mathbf{q}}^{(\mathrm{r})}S(\mathbf{q},\omega _{q})+(N_{-%
\mathbf{q}}^{(\mathrm{r})}+1)S(\mathbf{q},-\omega _{q})\right] ,  \label{e11}
\end{eqnarray}%
where $\omega _{q}=\sqrt{\alpha /\rho }q^{3/2}$ and $N_{\mathbf{q}}^{(%
\mathrm{r})}$ are the spectrum and distribution function of ripplons, $%
Q_{q}^{2}=\hbar q/2\rho \omega _{q}$, $\alpha $ and $\rho $ are the surface
tension and the liquid helium mass density, respectively. In the limit of
strong holding electric fields, the coupling parameter $V_{q}$ does not
depend on the wave number $V_{q}\simeq eE_{\bot }$. In general we have to
include the polarization term $V_{q}=e(E_{\bot }+E_{q})$, where $E_{q}$ has
a quite complicated dependence on the 2D wave number $q$~\cite%
{ShiMon74,PlaBen76}. The channel DSF $S(\mathbf{q},\omega _{q})$
depends on the average velocity $u_{\mathrm{av}}$, so that the sum
over the all wave vectors $\mathbf{q}$ entering Eq.~(\ref{e11}) is
not zero. Usually, the
determination of the relationship between $S(\mathbf{q},\omega _{q})$ and $%
u_{\mathrm{av}}$ is the most difficult part of the transport theory. For
highly correlated electrons ($\nu _{ee}\gg \nu _{\mathrm{eff}}$), there is a
great simplification because in the center-of-mass reference frame moving
along the channel such an electron system can be described by the
equilibrium DSF $S_{0}(\mathbf{q},\omega ).$ Therefore, in the laboratory
frame the frequency argument acquires the Doppler shift and $S(\mathbf{q}%
,\omega )\simeq S_{0}(\mathbf{q},\omega -q_{x}u_{\mathrm{av}})$~ \cite%
{MonTesWyd02}.

In the momentum-balance equation method~\cite{LeiTin84,CaiLeiTin85} $S_{0}(%
\mathbf{q},\omega -q_{x}u_{\mathrm{av}})$ is usually expanded in powers of $%
q_{x}u_{\mathrm{av}}$. Instead, the sum of Eq.~(\ref{e11}) can be rearranged
and presented in a more convenient form
\begin{eqnarray}
F_{\mathrm{scat}} &=&\frac{N_{e}}{\hbar S_{\mathrm{A}}}\sum_{\mathbf{q}%
}q_{x}V_{q}^{2}Q_{q}^{2}N_{\mathbf{q}}^{(\mathrm{r})}\hspace{1cm}  \nonumber
\\
&&\times \left[ 1-e^{\hbar q_{x}u_{\mathrm{av}}/T}\right] S_{0}(\mathbf{q}%
,\omega _{q}-q_{x}u_{\mathrm{av}}),  \label{e12}
\end{eqnarray}%
where we have used the properties of the equilibrium DSF $S_{0}(-\mathbf{q}%
,\omega )=S_{0}(\mathbf{q},\omega )$, and $S_{0}(\mathbf{q},-\omega )=\exp
(-\hbar \omega /T)S_{0}(\mathbf{q},\omega )$, which can be seen from Eq.~(%
\ref{e8}). This representation is convenient because the linear theory
result can be obtained by disregarding the Doppler correction $-q_{x}u_{%
\mathrm{av}}$ in the argument of the DSF leading to
\begin{equation}
m_{e}\nu _{\mathrm{eff}}^{(\mathrm{r})}(B)=\frac{1}{S_{\mathrm{A}}T}\sum_{%
\mathbf{q}}V_{q}^{2}Q_{q}^{2}q_{x}^{2}N_{q}^{(\mathrm{r})}S_{0}(\mathbf{q}%
,\omega _{q}).  \label{e13}
\end{equation}%
It is worthy of remark that the effective mass of electrons in the channel $%
m_{B}$ enters into the frictional force only by means of the channel DSF
(the free electron mass $m_{e}$ in the left-hand part of Eq.~(\ref{e13}) is
chosen as a convenient proportionality factor). In contrast to the isotropic
2D case, the DSF of the electron channel $S_{0}(\mathbf{q},\omega _{q})$
depends strongly on the direction of the wave vector $\mathbf{q}$.

The same expression for the effective collision frequency could be obtained
by means of the memory function formalism introduced by G\"{o}tze and W\"{o}%
lfle ~\cite{GotWol72}. In this approach the quantum conductivity
equation looks like an extension of the classical Drude formula,
in which the
imaginary part of the conductivity relaxation kernel [the memory function $%
M(\omega )$] plays the role of the effective collision frequency. It should
be noted that the approximation for the memory function frequently used in
quantum transport equations is actually a high-frequency approximation ($%
\omega \gg \nu $), even though it usually gives correct results in
the whole frequency range. Platzman \textit{et
al.}~\cite{PlaSimTzo77} were the first to apply this approach for
the analysis of the effect of electron correlations. The important
point is that the high-frequency approximation for the
conductivity relaxation kernel and the DC approach based on the
condition $\nu _{ee}\gg \nu $ employed here yield the same
equations for the effective collision frequency as it is in the
semiclassical kinetic equation method.

If the temperature is relatively high ($T\gtrsim 0.7\,\mathrm{K}$), we have
to consider the possibility of electron scattering by helium vapor atoms.
Even though the helium vapor atoms represent a sort of impurity-like
scatterers, we can disregard quantum localization effects because the
electron-electron collision rate is extremely high ($\nu _{ee}\gg \nu $). In
the quasi-elastic approximation the above described treatment leads to the
following correction to the effective collision frequency induced by vapor
atoms:
\begin{equation}
m_{e}\nu _{\mathrm{eff}}^{(\mathrm{a})}(B)=\frac{3n_{\mathrm{a}}V_{\mathrm{a}
}^{2}\gamma }{16TS_{\mathrm{A}}}\sum_{\mathbf{q}}q_{x}^{2}S_{0}(\mathbf{q}
,0),  \label{e14}
\end{equation}
where $n_{\mathrm{a}}$ is the density of vapor atoms, $\gamma $ is the
parameter describing the wave function of SE states $\psi _{1}(z)\propto
z\exp (-\gamma z)$, and $V_{\mathrm{a}}$ \ describes the interaction of a
free electron with a single helium atom $V(\mathbf{R}- \mathbf{R}_{a})=V_{%
\mathrm{a}}\delta (\mathbf{R}-\mathbf{R}_{a})$. Even though the interaction
parameter is usually written in the form $V_{\mathrm{a }}=2\pi \hbar
^{2}s_{0}/m_{e}$ containing the scattering length $s_{0}$ and $m_{e}$, the
effective mass of electrons in the channel $m_{B}$ appears only in the
channel DSF.

\section{Results and discussion}

It is instructive to consider the ultra-quantum limit $\hbar \Omega \gg T$.
In this case, the electron DSF can be approximated by the term with $%
n=n^{\prime }=0$:
\begin{equation}
S(\mathbf{q},0)\simeq \hbar \sqrt{\frac{\pi }{\varepsilon _{q_{x}}T}}\exp
\left( -\frac{q_{y}^{2}l_{y}^{2}}{2}-\frac{q_{x}^{2}l_{x}^{2}}{2}\right) ,
\label{e15}
\end{equation}
where the parameter $l_{x}$ was defined in Eq.~(\ref{e9b}). We consider only
the DC case, and disregard the frequency argument because $\hbar \omega
_{q}/T$ is quite small and the relevant parameter $\hbar \omega _{q}/\Gamma
^{\ast }$ is small if the $B$ is not too high. In this approximation Eq.~(%
\ref{e13}) turns out to be

\begin{eqnarray}
m_{e}\nu _{\mathrm{eff}}^{(\mathrm{r})}=\frac{2m_{B}^{1/2}}{(2\pi
)^{3/2}\alpha T^{1/2}}\int\limits_{0}^{\infty
}dq_{x}q_{x}e^{-q_{x}^{2}l_{x}^{2}/2}  \nonumber \\
\times \int\limits_{0}^{\infty }dq_{y}\left[ V_{q}^{2}/%
\left(q_{x}^{2}+q_{y}^{2}\right) \right] e^{-q_{y}^{2}l_{y}^{2}/2}. \hspace{%
1cm}  \label{e16}
\end{eqnarray}

We point out that, in contrast with the pure 2D case where $\nu _{\mathrm{eff%
}}$ and $\sigma _{xx}$ are proportional to $1/\sqrt{T}$, the effective
collision frequency [Eq.~(\ref{e16})] is finite in the formal limit $%
T\rightarrow 0$ because $l_{x}^{2} $ contains the temperature dependent term
according to Eq.~(\ref{e9b}). For a pure 2DES ($l_{x}=l_{y}=l_{B}$), Eq.~(%
\ref{e16}) also results in $\nu _{\mathrm{eff}}^{(\mathrm{r})}\propto 1/%
\sqrt{T}$. For the electron channel, this behavior is limited to the
temperature range where the temperature dependent term of $l_{x}^{2}$ is
small.

In general $V_{q}$ is a very complicated function of the 2D wave number $q$,
and Eq.~(\ref{e16}) should be evaluated numerically for given channel
parameters. If the electron channel is formed by applying a strong holding
electric field $E_{\bot }$, i.e. $E_{\bot }\gg E_{q}$ we can use the
approximation $V_{q}\simeq eE_{\bot }$. Then, the effective collision
frequency of the electron channel can be found analytically as
\begin{eqnarray}
m_{e}\nu _{\mathrm{eff}}(B)=\frac{ 2m_{e}\nu ^{(0)}(B)}{\pi \sqrt{1-4T/\hbar
\Omega }}\hspace{1.5cm}  \nonumber \\
\times\arctan \left[ \left( \hbar \omega _{0}^{2}/4\Omega T\right) ^{1/2}%
\sqrt{1-4T/\hbar \Omega }\right] ,  \label{e18}
\end{eqnarray}
where
\begin{equation}
m_{e}\nu ^{(0)}(B)=\left( \frac{m_{B}}{m_{e}}\right) ^{1/2}\frac{(eE_{\bot
})^{2}}{2\alpha l_{y}^{2}\omega _{0}}.  \label{e19}
\end{equation}
We can see that only the increase of the effective mass described by $%
m_{B}=m_{e}\Omega ^{2}/\omega _{0}^{2}$ cannot explain the whole magnetic
field dependence of the electron mobility. The field dependence of the
effective collision frequency is determined by the interplay of the mass
enhancement and the field dependencies of $l_{x}(B)$ and $l_{y}(B)$. For
example, an additional increase of the channel magnetoresistance appears
because of the factor $1/l_{y}^{2}\propto \Omega $ which comes from the
scattering matrix elements. Recalling the definitions of $m_{B}$ and $%
l_{y}^{2}$ given above we can find that the the low temperature limit $\nu
^{(0)}(B)$ of the effective collision frequency $\nu _{ \mathrm{eff}}(B)$
increases with $B$ as
\begin{equation}
\nu ^{(0)}(B)=\left( 1+\frac{\omega _{c}^{2}}{\omega _{0}^{2}}\right) \frac{%
(eE_{\bot })^{2}}{2\alpha \hbar }.  \label{e20}
\end{equation}
For zero magnetic field, $\nu ^{(0)}(B)$ tends to $\nu _{0}=(eE_{\bot
})^{2}/2\alpha \hbar $.

\begin{figure}[tbp]
\includegraphics*[width=0.8\linewidth]{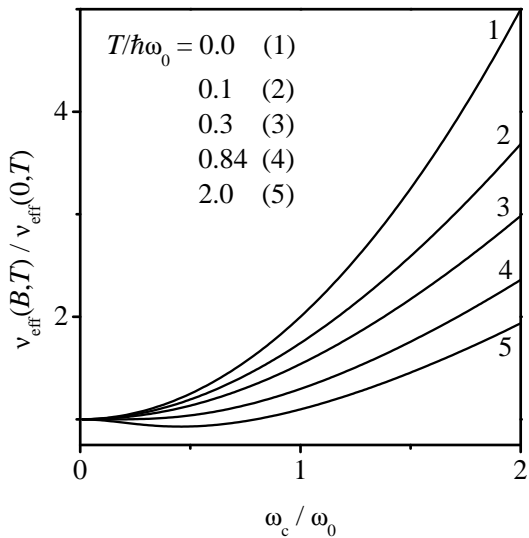}
\caption{Magnetic field dependence of the electron effective
collision frequency due to ripplons given in
Eq.~(\protect\ref{e18}) for different temperatures. } \label{f3}
\end{figure}

Under the condition $4T\Omega /\hbar \omega _{0}^{2}\ll 1$, Eq.~\ (\ref{e18}%
) gives the following asymptote for the channel mobility
\begin{equation}
\mu \simeq \frac{2\alpha \hbar }{m_{e}eE_{\bot }^{2}}\left( \frac{\omega _{0}%
}{\Omega }\right) ^{2}\left[ 1+\frac{4}{\pi }\left( \frac{T\Omega }{\hbar
\omega _{0}^{2}}\right) ^{1/2}\right] .  \label{e21}
\end{equation}%
For $B=0$, this equation reproduces the result found previously in Ref.~\
\cite{SokHaiStu95}. It is remarkable that the zero-temperature term of Eq.~(%
\ref{e21}), as a function of $B$, agrees with the formal mass replacement $%
m_{e}\rightarrow m_{B}=m_{e}\Omega ^{2}/\omega _{0}^{2}$ in the mobility
equation $\mu =2\alpha \hbar /m_{e}eE_{\bot }^{2}$ obtained for electron in
the channel at $B=0$. We also note that the temperature correction increases
with $B$ and affects strongly the magnetoresistivity of the electron
channel. Moreover, at high magnetic fields the condition $4T\Omega /\hbar
\omega _{0}^{2}\ll 1$ breaks down and we have to use the more general
expression given in Eq.~(\ref{e18}). For different temperatures the magnetic
field dependence of this expression is shown in Fig.~\ \ref{f3} using the
normalized units $T/\hbar \omega _{0}$ and $\omega _{c}/\omega _{0}$. We can
see that even at relatively low temperatures ($T/\hbar \omega _{0}=0.1$), $%
\nu _{\mathrm{eff}}(B)/\nu _{0}$ deviates strongly from the zero-temperature
asymptote. It is remarkable that, for $T/\hbar \omega _{0}>0.84$, we observe
a negative magnetoresistance in the region of low fields. The origin of this
unusual behavior is the nonmonotonous field dependence of $l_{x}^{2}(B)$
discussed above and shown in Fig.~\ref{f2}. Still, at such high
temperatures, we cannot neglect the terms with $n,n^{\prime }>0$ in the
equation for the channel DSF.

\begin{figure}[tbp]
\includegraphics*[width=0.8\linewidth]{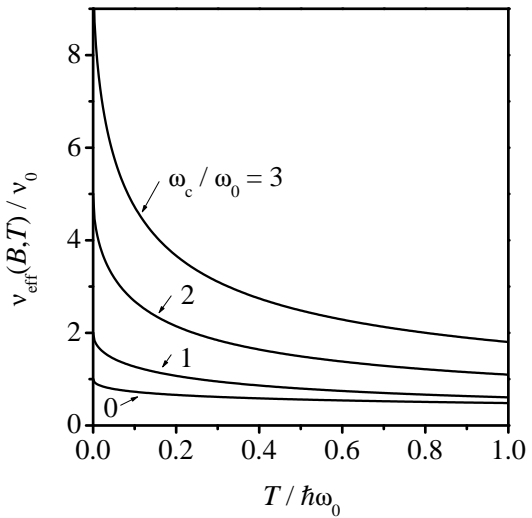}
\caption{Temperature dependence of the electron effective
collision frequency due to ripplons given in
Eq.~(\protect\ref{e18}) for different magnetic fields. }
\label{f4}
\end{figure}

Figure~\ref{f4} shows how the magnetic field affects the temperature
dependence of the resistivity of the electrons in the channel. For zero
magnetic field (line 0), the temperature dependence is relatively weak. The
magnetic field makes the temperature dependence sharper in the
low-temperature range (lines representing $\omega _{c}/\omega _{0}= 1,2$,
and $3$), which is consistent with the singular nature of the 2DES subject
to the normal magnetic field because at high fields ($l_{y}\rightarrow l_{B}$%
) the channel becomes effectively a quasi-2DES. Physically, the effective
collision frequency increases due to the multiple electron scattering. The
number of multiple scattering events is limited by the electron velocity
which decreases strongly with $B$.

At high fields and low temperatures the effective broadening
$\Gamma ^{\ast }\approx \sqrt{2}eE_{\mathrm{ch}}l_{B}$ of the
channel DSF  decreases as $%
\Gamma ^{\ast }\propto \sqrt{T/B}$. In this regime the electron
channel becomes similar to the 2DES in which the broadening
parameter $%
\Gamma _{n,n^{\prime }}$ of the DSF is determined by the
interaction with scatterers.
In the self-consistent Born approximation $\Gamma _{n,n^{\prime}}=\sqrt{%
(\Gamma _{n}^{2}+\Gamma _{n^{\prime }}^{2})/2}$~\cite{MonTesWyd02}, where $%
\Gamma _{n}$ is the collision broadening of the Landau levels. The collision
broadening usually increases with $B$. Therefore, the above results are
valid under the condition $\sqrt{2}eE_{\mathrm{ch}}l_{B}>\Gamma _{0}$. For
the electron-ripplon interaction, the collision broadening has the same
temperature dependence ($\Gamma _{0}\propto \sqrt{T}$), and the breakdown of
the channel magnetotransport equations occurs when the magnetic field is
increased beyond the value given by the above condition.

\begin{figure}[tbp]
\includegraphics*[width=0.8\linewidth]{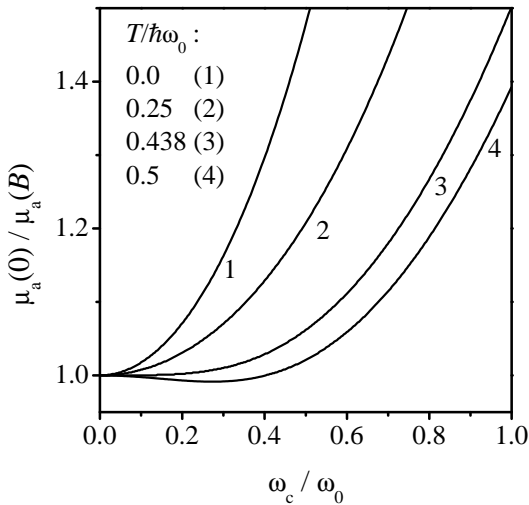}
\caption{The inverse mobility of electrons in the channel vs
magnetic field for different temperatures under the condition that
electrons are predominantly scattered by short-range impurity like
scatterers (vapor atoms). } \label{f5}
\end{figure}

For electron scattering by vapor atoms in the ultra-quantum conditions, Eq.~(%
\ref{e14}) can be evaluated analytically
\begin{equation}
m_{e}\nu _{\mathrm{eff}}^{(\mathrm{a})}(B)=\frac{3n_{\mathrm{a}}V_{\mathrm{a}
}^{2}\gamma m_{B}^{1/2}}{16\pi T^{3/2}l_{y}l_{x}^{2}}.  \label{e22}
\end{equation}
Using the dimensionless units $\Delta _{c}=\omega _{c}/\omega _{0}$ and $%
\tau =T/\hbar \omega _{0}$, the corresponding mobility of the electrons in
the channel can be written as
\begin{equation}
\frac{\mu _{\mathrm{a}}(B)}{\mu _{\mathrm{a}}(0)}=\frac{1+4\tau \Delta
_{c}^{2}/\sqrt{1+\Delta _{c}^{2}}}{(1+\Delta _{c}^{2})^{7/4}}.  \label{e23}
\end{equation}
In this case the negative magnetoresistance originating from the field
dependence of $l_{x}^{2}(B)$ becomes more prominent for such short-range
impurity-like scatterers, as shown in Fig.~\ref{f5}. It should be noted that
the negative magnetoresistance of the electrons occupying the lowest level ($%
n=0$) resembles that observed for quantum localization effects in
nondegenerate 2DES ~\cite{Ada97,KarHerMat00} under the condition
$\nu _{ee}\ll \nu $, although in the transport regime $\nu
_{ee}\gg \nu $ considered here quantum localization effects can be
disregarded. However if the parameter $\tau $ is not small, the
other terms of the DSF ($n,n^{\prime }>0$) are important and the
system behaves like the 2DES exhibiting positive magnetoresistance
at $\nu _{ee}\gtrsim \nu $. In this case we have to perform
numerical calculations.

\begin{figure}[tbp]
\includegraphics*[width=0.8\linewidth]{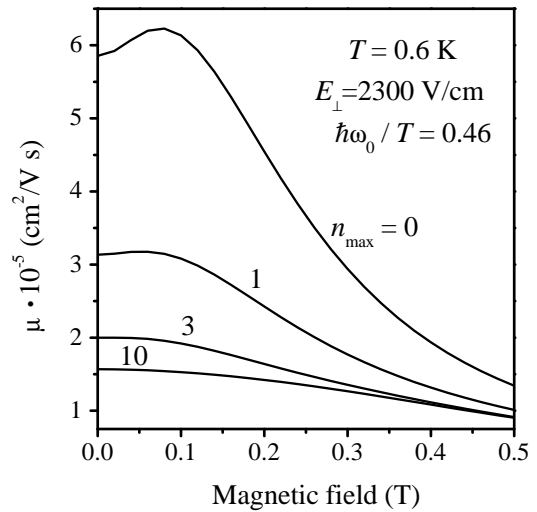}
\caption{Mobility of electrons in the channel vs magnetic field
under the
condition that channel levels with $n>0$ are occupied ($T/\hbar \protect%
\omega _{0}\simeq 2.2$). The succession of solid curves show how the
inclusion of higher channel levels with $n,n^{\prime}\leq n_{\max }$
diminishes the negative magnetoresistance in the region of weak fields. }
\label{f6}
\end{figure}

Typical field dependencies of the channel mobility evaluated numerically
using Eqs.~(\ref{e13}) and (\ref{e14}) for different $n_{\max }=n_{\max
}^{\prime }$ which restrict the sums over $n$ and $n^{\prime }$ are depicted
in Fig.~\ref{f6}. For $\hbar \omega _{0}/T=0.46$, the electrons
predominantly occupy the high energy levels. We observe clearly the negative
magnetoresistance only when $n_{\max }=0$ and $n_{\max }=1$. The inclusion
of high levels suppresses the negative magnetoresistance of the electrons in
the channel, and in the limit $n_{\max }\gg 1$ the electron channel shows
only a positive magnetoresistance because, as expected, it becomes
effectively a quasi-2DES.

\begin{figure}[tbp]
\includegraphics*[width=0.8\linewidth]{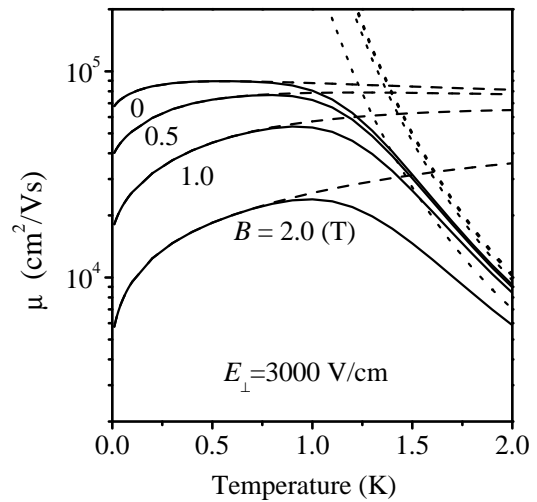}
\caption{The mobility of electrons in the channel vs temperature
evaluated including all channel levels: \textit{solid curves} show
the effect of both scattering mechanisms, while other curves
represent electron scattering by ripplons (\textit{dashed curves})
and vapor atoms (\textit{dotted curves}) solely.} \label{f7}
\end{figure}

For fixed values of the magnetic field, typical temperature dependencies of
the channel mobility evaluated numerically including high levels ($%
n,n^{\prime }\gg 1$) are shown in Fig.~\ref{f7}. The solid lines of the
figure represent the contributions from both scattering mechanisms which
interplay at $T\approx 1\, \mathrm{K}$, while the dotted and dashed lines
show the separate contributions from scattering by vapor atoms and ripplons
correspondingly. For $T<1\, \mathrm{K}$ electrons are predominantly
scattered by ripplons. In this regime at high magnetic fields, there is a
range where the electron mobility $\mu \propto \sqrt{T}$. Nevertheless, at
lower temperatures $\mu $ approaches a finite value which decreases strongly
with $B$.

It should be noted that in the limit of high $B$, the inelastic effect of
the electron-ripplon interaction cannot be disregarded. In this case the
expression in the integrand appearing in Eq.~(\ref{e16}) contains an
additional exponential factor $\exp [-(\hbar \omega _{q}/\Gamma ^{\ast
})^{2}]$ because of the finite frequency argument of the electron DSF $S_{0}(%
\mathbf{q},\omega _{q})$. Typical ripplon wave numbers and frequencies
involved in scattering events increase strongly with $B$, while the
effective broadening of the DSF of the electron channel $\Gamma ^{\ast }=2%
\sqrt{\varepsilon _{q_{x}}T}$ decreases (it decreases also by lowering $T$).
Eventually, the energy exchange $\hbar \omega _{q}$ at a collision becomes
comparable with $\Gamma ^{\ast }$ and the effective collision frequency
decreases with cooling the system.

For the pure 2DES realized on the surface of liquid helium such an inelastic
effect was discussed in Ref.~\cite{MonItoShi97}. It was shown that for $%
T>0.1\, \mathrm{K}$, the decrease of $\nu _{\mathrm{eff}}^{(\mathrm{r})}$
due to inelastic effects becomes important only if $B>2\, \mathrm{T}$
(actually it was observed at $B=6.4\, \mathrm{T}$). In this paper we
consider lower magnetic fields to keep $\omega _{c}$ comparable with the
confinement frequency $\omega _{0}$. Additionally, $\Gamma ^{\ast }$ is
assumed to be larger than the collision broadening of the Landau levels. In
this range inelastic effects on the channel magnetotransport can be
disregarded.

In conclusion, we have shown that nondegenerate quantum wires may constitute
a remarkable laboratory for testing the quantum transport theory. In
contrast with classical wires, they exhibit strong magnetoresistance (the
channel mobility decreases with the field intensity) which is a result of
the interplay between the mass-enhancement of channel carriers $m_{B}$ and
the magnetic-field-induced increase of the momenta exchange ($\hbar q_{x}$
and $\hbar q_{y}$) in scattering events. For highly correlated electrons ($%
\nu _{ee}\gg \nu $), we have obtained the general relation between the
channel mobility and the dynamical structure factor of such anisotropic
electron system. Evaluations performed for particular cases indicate that
the effect of a normal magnetic field on the channel mobility is very strong
for conducting channels formed on the surface of liquid helium under usual
experimental conditions.

\acknowledgments This work was supported by the Brazilian funding
agencies FAPESP and CNPq.


\begin{thebibliography}{9}
\bibitem{And97} {\it Two-Dimensional Electron Systems
on Helium and other Substrates}, edited by E. Andrei (Kluwer
academic publishers; Dordrecht, Boston, London, 1997).

\bibitem{MonTesWyd02} Yu.P. Monarkha, E. Teske, and P. Wyder, Physics
Reports \textbf{370}, issue 1, pages 1-61 (2002).

\bibitem{PetSchLea94}  P.J.M. Peters, P. Scheuzger, M.J. Lea, Yu.P.
Monarkha, P.K.H. Sommerfeld, and R.W. van der Heijden, Phys. Rev.
B \textbf{50,} 11570 (1994).

\bibitem{LeaFozRic94}  M.J. Lea, P. Fozooni, P.J. Richardson, and A.
Blackburn, Phys. Rev. Lett., \textbf{73}, 1142 (1994).

\bibitem{MonTesWyd99}  Yu.P. Monarkha, E. Teske, and P. Wyder, Phys.
Rev.B \textbf{59}, 14884 (1999).

\bibitem{KirMonKov93} O.I. Kirichek, Yu.P. Monarkha, Yu.Z. Kovdrya, and V.N.
Grigor'ev, Low Temp. Phys. \textbf{19}, 323 (1993) [Fiz. Nizk.
Temp. \textbf{19}, 458 (1993)].

\bibitem{ValHei98}  A.M.C. Valkering, and R.W. van der Heijden, Physica
B \textbf{249-251}, 652 (1998).

\bibitem{SokStu95} S.S. Sokolov and N. Studart, Phys. Rev.
B \textbf{51}, 2640 (1995).

\bibitem{Gir98} S.M. Girvin, in {\it Topological aspects of low dimensional
systems}, ed by A. Comtet, T. Jolic\ae ur, S. Ouvry, F. David (EDP
sciences and Springer-Verlag, Berlin 1998).

\bibitem{GlaDotFoz01}  P. Glasson, V. Dotsenko, P. Fozooni, M.J. Lea, W. Bailey,
G. Papageorgiou, S.E. Andersen, and A. Kristensen, Phys. Rev.
Lett. \textbf{87}, 176802 (2001).

\bibitem{StuSok97} N. Studart and S.S. Sokolov: in \textit{Two-dimensional
electron systems on helium and other Cryogenic Substrates}, edited
by E. Andrei (Kluwer academic publishers; Dordrecht, Boston,
London, 1997).

\bibitem{GotHaj78}  W. G\"{o}tze and J. Hajdu, J. Phys. C \textbf{11}, 3993
(1978).

\bibitem{AndFowSte82}  T. Ando, A.B. Fowler, and F. Stern, Rev. Mod. Phys.
\textbf{54}, 437 (1982).

\bibitem{LeaFozKri97}  M.J. Lea, P. Fozooni, A. Kristensen, P.J.
Richardson, K. Djerfi, M.I. Dykman, C. Fang-Yen, and A. Blackburn,
Phys. Rev. B \textbf{55}, 16280 (1997).

\bibitem{VilMon89}  Yu.M. Vil'k and Yu.P. Monarkha, Sov. J. Low Temp.
Phys. \textbf{15}, 131 (1989) [Fiz. Nizk. Temp. \textbf{15}, 235
(1989)].

\bibitem{ShiMon74}  V.B. Shikin and Yu.P. Monarkha, J. Low Temp. Phys.
\textbf{16}, 193 (1974).

\bibitem{PlaBen76}  P.M. Platzman, and G. Beni, Phys. Rev. Lett.
\textbf{36}, 626 (1976).

\bibitem{LeiTin84}  X.L. Lei and C.S. Ting: Phys. Rev. B \textbf{30}, 4809
(1984).

\bibitem{CaiLeiTin85}  W. Cai, X.L. Lei, and C.S. Ting, Phys. Rev.
B \textbf{31}, 4070 (1985).

\bibitem{GotWol72}  W. G\"{o}tze and P. W\"{o}lfle, Phys. Rev. B \textbf{6},
1226 (1972).

\bibitem{PlaSimTzo77}  P.M. Platzman, A.L. Simons, and N. Tzoar, Phys. Rev.
B \textbf{16}, 2023 (1977)

\bibitem{SokHaiStu95}  S.S. Sokolov, G.-Q. Hai, and N. Studart, Phys. Rev. B
\textbf{51}, 5977 (1995).

\bibitem{Ada97} P.W. Adams: in {\it Two-dimensional electron systems on helium
and other Cryogenic Substrates}, edited by E. Andrei (Kluwer
academic publishers; Dordrecht, Boston, London, 1997).

\bibitem{KarHerMat00} I. Karakurt, D. Herman, H. Mathur, and A.J.
Dahm, Phys. Rev. Lett. \textbf{85}, 1072 (2000).

\bibitem{MonItoShi97}  Yu.P. Monarkha, S. Ito, K. Shirahama, and K.
Kono, Phys. Rev. Lett. \textbf{78}, 2445 (1997).

\end{thebibliography}
\end{document}